# Compact laser-driven plasma X-ray source for time-resolved diffraction, spectroscopy, and imaging experiments at ELI Beamlines

Y. Pulnova[a,b], T. Parkman[a,c], B. Angelov[a], I. Baranova[a,b], A. Zymaková[a], S. Cipiccia[d], L. Fardin[d], B.A. Yorke[e], R. Antipenkov[a], D. Peceli[a], O. Hort[a], D.D. Mai[a], J. Andreasson[a], and J. Nejdl[a,c*] a: Extreme Light Infrastructure ERIC, ELI Beamlines facility, Za Radnicí 835, 252 41 Dolní Břežany, Czech Republic

b: Charles University, Faculty of Mathematics and Physics, 121 16 Prague, Czech Republic

c: Czech Technical University in Prague, Jugoslávských partyzánů 1580/3, 160 00 Prague, Czech Republic

d: Department of Medical Physics and Biomedical Engineering, University College London, 2 Malet Pl, London, WC1E 7JE, UK

e: School of Chemistry, University of Leeds, Woodhouse Lane, Leeds, LS2 9JT, UK

*Correspondence email: jaroslav.nejdl@eli-beams.eu

**Synopsis**  A laser-driven plasma X-ray source with sub-ps pulses at 1 kHz repetition rate for various time-resolved experiments has been commissioned at ELI Beamlines. The paper features a comprehensive overview of driving laser parameters and X-ray beam characteristics and outlines possible applications of the source.

**Abstract**  Experimentally measured characteristics of a kHz laser-driven Cu plasma X-ray source that was recently commissioned at ELI Beamlines facility are reported. The source can be driven either by an in-house developed high contrast sub-20 fs near-infrared TW laser based on optical parametric chirped-pulse amplification technology, or by a more conventional Ti:Sapphire laser delivering 12 mJ, 45 fs pulses. The X-ray source parameters obtained with the two driving lasers are compared. Measured photon flux of the order up to $10^{12}$ $K_\alpha$ photons/$4\pi$/s is reported. Furthermore, experimental platforms for ultrafast X-ray diffraction and X-ray absorption and/or emission spectroscopy based on the reported source are described.

## 1. Introduction

Laser-driven plasma X-ray sources (PXS) hold a niche in studying ultrafast phenomena thanks to their ability to provide pulses as short as hundreds of femtoseconds. Moreover, their compact size makes them favorable for small-scale laboratories which enhances their availability to a broad scientific community. X-ray pulses of such short duration are otherwise achievable only by the synchrotron slicing technique (Prigent *et al.*, 2013), LINAC sources (Enquist *et al.*, 2018) or at XFEL facilities (Altarelli, 2011; Schietinger, 2018) and allow time resolved studies in material science (Mattern *et al.*, 2021) molecular dynamics (Freyer *et al.*, 2013) and structural biology (Khakurel *et al.*, 2024). The two main application branches are ultrafast time-resolved X-ray diffraction (XRD), using the characteristic line emission of the target material, and X-ray absorption and emission spectroscopy (XAES) that benefits mainly from the continuous Bremsstrahlung emission of hot electrons' collisions in the target.

Laser-driven plasma sources for X-ray generation have been demonstrated across solid, liquid, and gas targets with a diverse range of flux and source sizes. Sources with solid targets, such as Cu, Mo, and Ti (Holtz *et al.*, 2017; Afshari *et al.*, 2020; Li *et al.*, 2017; Azamoum *et al.*, 2018; Rousse, Audebert *et al.*, 1994), reach photon fluxes up to $10^{12}$ photons/s/4π with source sizes varying from a few microns to several tens of microns. Solid target sources, used together with high power lasers, presently produce the highest brightness among PXSs but are prone to long-term flux decline due to target deformation and damage, as well as higher operational cost due to the consumable target. Liquid sources, such as Ga and Hg jets (Reich *et al.*, 2007; Zhavoronkov *et al.*, 2004; Ivanov *et al.*, 2011), show photon fluxes around $10^9$ photons/s/4π and are mostly used with lower peak power lasers. These sources hold the potential of long-term, low-maintenance operation by recycling the target metal. However, technical challenges, such as maintaining a homogeneous target temperature and effective debris mitigation, are yet to be solved, in particular when increasing the driving laser power. Gas-based sources, such as argon (Chen *et al.*, 2007, 2010), allow the use of much more powerful laser pulses due to inherent debris-free operation, allowing for the fluxes of up to $10^{11}$ photons/s/4π. Nevertheless, they come with a much larger source size, up to 100 microns. This prevents their use for applications requiring some degree of spatial coherence.

For solid target PXS, the main challenge is maintaining its operation stability. The shot-to-shot instabilities are mainly caused by mechanical vibrations of the target and fluctuations in the laser pointing. A decrease in X-ray flux on timescale of several hours is usually observed. This is due to the debris accumulation affecting the mechanical stability of dynamical parts (e.g. guiding bearings), or, in some setups, coating the IR focusing optics and reducing their efficiency. Several normalization strategies for data processing applicable to both types of instability have been suggested (Schick *et al.*, 2012; Zhang *et al.*, 2014), commonly requiring the introduction of the second X-ray detector. Further efforts are focused on automation control (Zhao *et al.*, 2022) and mechanical design improvements.

The choice of the driving laser also affects the performance of the X-ray source. Plasma physics models and experimental results indicate that increasing the driver wavelength leads to higher X-ray photon flux

(Weisshaupt *et al.*, 2014; Koç *et al.*, 2021), accompanied by a relative increase of the X-ray source size. The presence of the ns pre-pulse decreases the X-ray yield (Eder *et al.*, 2000), whereas the pulse duration determines the maximum temporal resolution for the pump-probe experiments. Plasma X-ray sources require a stroboscopic approach for achieving appropriate signal-to-noise ratio. The total exposure time is then highly dependent on the available repetition rate. For the solid-target PXS, it ranges from 10 Hz with nowadays established maximum of 10 kHz. Target Z-number also affects the X-ray yield of a source with given driving laser parameters (Rousse, Audebert *et al.*, 1994).

ELI Beamlines is a part of The Extreme Light Infrastructure ERIC that offers a wide range of high-power lasers, laser-driven sources of short-wavelength radiation, and beams of accelerated particles (Rus *et al.*, 2013). A variety of end-stations enable numerous ultrafast experiments relevant to many fields, such as atomic, molecular, and optical physics, material and life sciences, plasma physics, etc. A copper tape PXS is offered to ELI users as a source with unique set of parameters for time-resolved X-ray diffraction and spectroscopy experiments, as well as for X-ray imaging applications.

In this paper, we report on the Cu tape target PXS at ELI Beamlines, together with two setups for applications in time-resolved X-ray diffraction and spectroscopy. The article is organized in the following way: Section 2 is devoted to the physics of X-ray generation in a PXS. Section 3 describes the laser systems available for driving our X-ray source and reports on measurement of their characteristics. Section 4.1 describes the target system that provides a renewable solid Cu target. In section 4.2, we present the measured parameters of the PXS. In section 5.1, the time-resolved experimental station for ultrafast diffraction (TREX) is presented, and Section 5.2 offers an outlook of the X-ray absorption and emission (XAES) end-station. We provide a discussion and a summary in Section 6.

## 2. Theoretical Background: X-ray Generation Mechanisms in laser plasma

An important milestone for compact laser-driven plasma X-ray sources was enabled by invention of the chirped pulse amplification method that introduced the high-peak power sub-ps laser technology (Strickland & Mourou, 1985). Consequently, the dominant laser absorption mechanism shifted from the resonant absorption with longer pulses (Forslund *et al.*, 1977) to the vacuum heating (Brunel, 1987) with short ones, provided that the laser pulse has sufficient contrast.

In the long-pulse regime, the heated part of the target undergoes expansion and the region with most absorption occurs at lower plasma density *n*. The plasma frequency in this region given by $\omega_p = \sqrt{e^2 n / m_e \varepsilon_0}$, where $e$ and $m_e$ are the electron charge and mass, respectively, and $\varepsilon_0$ is vacuum permittivity, is close to the laser frequency. The laser energy is absorbed by collisions of plasma electrons that oscillate in the electric field of the laser. Given that plasma expands with the speed of sound $c_s$, this happens when the laser pulse length is longer than 100 fs (Murnane *et al.*, 1991).

For short pulses interacting with solid targets, the created plasma density profile is much steeper. The laser pulse terminates before the required density for laser propagation is achieved through plasma expansion. Moreover, during laser illumination, the applied electric field can pull electrons off the surface of thin plasma layer out of the target. Following laser half-cycle, electrons are re-accelerated back to the target, delivering the acquired energy to the bulk through collisions with neutral atoms.

In both cases, X-ray radiation is produced by energetic electrons, which collide with ions and neutral atoms. These collisions can be either inelastic, ionizing the target atoms by creating vacancies in the inner electronic shells, which allows electrons from upper shells to decay and emit characteristic X-ray fluorescence photons, or they can be elastic and produce continuous bremsstrahlung radiation.

### 3. Driving kHz lasers available at ELI Beamlines

The Cu-tape X-ray source at ELI Beamlines facility can be driven by one of the kHz laser systems operated in experimental hall E1: the in-house developed laser L1 ALLEGRA or the commercial LEGEND Elite Duo laser (from Coherent). While LEGEND (12 mJ, 40 fs) is a chirped pulse amplification (CPA) system with a Ti:Sapphire oscillator and a regenerative amplifier followed by a one-pass booster amplifier, L1 ALLEGRA utilizes optical parametric chirped pulse amplification (OPCPA) for achieving ultrashort pulses with 15 fs duration and energy up to 50 mJ in a pulse (Antipenkov *et al.*, 2021). The L1 ALLEGRA nanosecond contrast corresponds to the total gain of all amplification stages, which is more than $10^7$, whereas for LEGEND it is as low as $10^3$, as seen at the oscilloscope. Both systems operate at similar central wavelengths (see Figure 1 *left*), allowing for convenient switching between them using the same beam transport optics.

The picosecond temporal contrast of both driving lasers L1 ALLEGRA and LEGEND is shown in Figure 1 *(right)*. Both measurements were performed using a SEQUOIA 800 cross-correlator from Amplitude Systems. The background noise and the limit of detection were determined to be around the $10^{-10}$ level by blocking each arm of the cross-correlator separately. For L1 ALLEGRA, the contrast was measured using sub-20 fs pulse duration with approximately 500 µJ energy, revealing a 10-ps contrast base level of around $10^{-9}$. Presumably, there are two weak pre-pulses as close as 2 ps and 1.5 ps before the main pulse with relative intensity $10^{-6}$. A peak observed at -3.2 ps is identified as a ghost peak, because its amplitude is approximately equal to the square of the amplitude of the post-pulse peak at +3.2 ps. In third-order correlation measurements, such ghost pre-pulse peaks arise due to the mixing of the second harmonic of the post-pulse with the main pulse at fundamental frequency (Tavella *et al.*, 2005; Chen *et al.*, 2024). For

LEGEND, the contrast was measured using 45-fs pulse with approximately 500 µJ energy, revealing a 10-ps contrast base level of around $10^{-7}$. However, there is a pre-pulse observed at -14 ps with intensity $10^{-4}$ and a ps-long pedestal with intensity of $10^{-2}$.

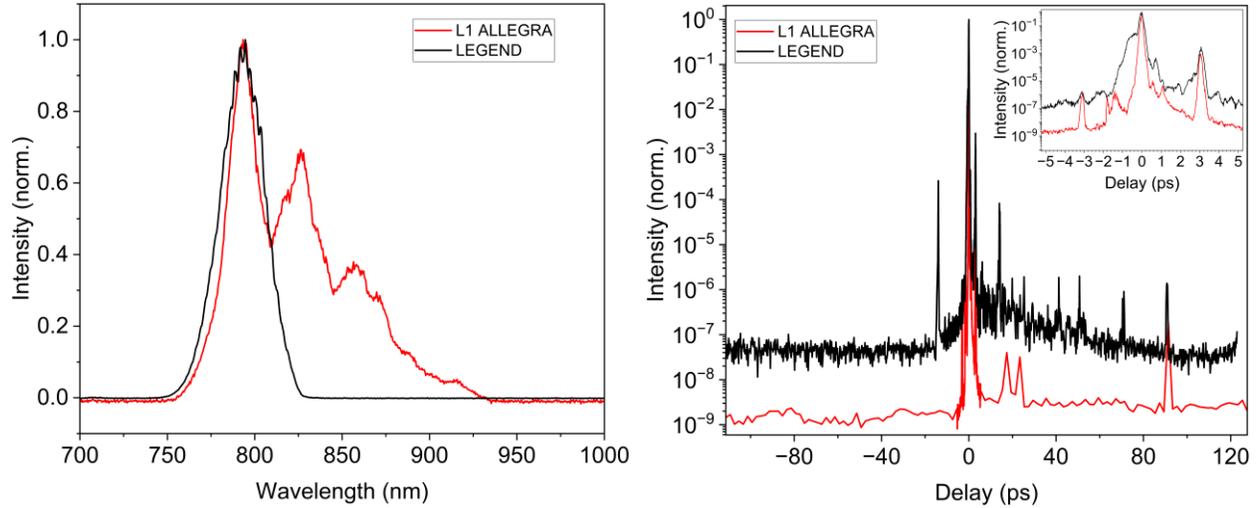

**Figure 1** *(Left)* Measured spectra of 1 kHz lasers available for driving the Cu-tape PXS: L1 ALLEGRA with central wavelength of 820 nm and LEGEND with central wavelength of 795 nm. *(Right)* Picosecond temporal contrast of the driving lasers L1 ALLEGRA and LEGEND, highlighting a pre-pulse with intensity $10^{-4}$ at -14 ps for LEGEND and a ghost peak at -3.2 ps for both lasers.

## 4. Plasma X-ray Source

### 4.1. Technical description

The PXS is situated in a radiation shielded hutch, with a footprint of 2.4 x 3.6 m. Its walls have a thickness of 6 mm and consist of 2 mm of lead sealed from both sides with 2 mm of steel. The layout of the interior of Cu-tape plasma X-ray source is illustrated in Figure 2. The chamber is shielded with a lead-steel housing, and it consists of three connected vacuum chambers operating at a pressure of $10^{-4}$ mbar. The upper and lower chambers contain three spools each that regulate the motion of the copper tape and two mylar bands. Spools are connected with driving motors outside of the chamber with 90-degree joints. The interaction chamber, the smallest one, is where the laser is focused onto the Cu tape target using a 2-inch gold-coated 90 degree off-axis parabola (OAP) with focal length of 76.2 mm. The OAP is installed on a motorized kinematic mount for convenient alignment. The copper tape is guided by a set of bearings which define the laser incidence angle of 23° from the target normal. The p-polarization of the laser pulse ensures efficient X-ray generation. The laser is focused onto the target through a 1 mm thick AR-coated vacuum window. The X-ray burst is emitted isotropically, but it exits the interaction chamber in beams that are defined by two exit slits enclosed by a 50 µm thick polyimide tape. The main beam dedicated to applications is situated along the laser propagation axis, the second beam is pointing upwards at 45° angle and it is used for diagnostic measurements and source alignment (see Section 4.2 for details).

The PXS at ELI Beamlines uses a 20 mm wide, 20 μm thick copper tape target. The velocity of the tape can be adjusted based on the plasma spot's size so that the surface damage of neighboring interaction spots does not overlap, and each laser pulse impacts the renewed target. The turning points of the copper tape are monitored and controlled by capacity sensors that are activated when the growing spool reaches its expected maximum size. The motion of the Cu-tape is stopped at this point. For starting the new line, the PXS chamber shifts horizontally, perpendicular to the laser axis, while the OAP is decoupled from this motion, therefore the focal spot remains fixed in space. This approach ensures that the X-ray source stays at the same position with respect to the downstream X-ray optics and no realignment is needed when the new line is started. An ultrafast Uniblitz laser shutter (80 ms full cycle) is automatically closed during the line-changing sequence to prevent the tape from being cut by the laser during the chamber shift.

As a large amount of copper dust debris is produced from the laser impact, two mylar foils are used to protect the vacuum windows. Moving tapes continuously remove the debris and their motion is also regulated by the capacitive sensors. On activation of the capacitive sensor, the tape automatically reverses its spooling direction. The interaction chamber is separated from the rest of the vacuum system only by thin slits for the tapes to reduce the debris in the upper and lower vacuum chambers and protect the spooling system.

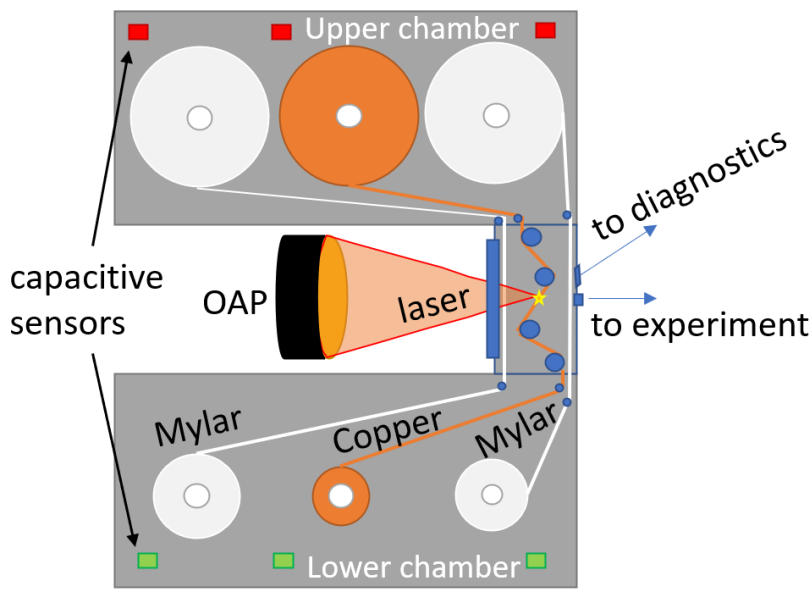

**Figure 2** The layout of the PXS. The U-shaped vacuum chamber contains the spools of the copper tape as well as shielding Mylar tapes that protect the entrance and exit windows from copper dust. The off-axis parabola focuses the laser onto the copper tape through the AR-coated vacuum window. The laser interaction point is marked by a star. The X-rays are emitted in $4\pi$ solid angle. Radiation exits through the two Kapton-sealed slits, towards the experiment and diagnostics, respectively. The capacitive sensors are marked with red rectangles, if switched by the spool size, and green, if not activated.

**4.2. X-ray source parameters**

A typical X-ray spectrum of the Cu PXS (shown in Figure 3) consists of two pronounced characteristic lines $K_\alpha$ at 8.04 keV and $K_\beta$ at 8.91 keV and a broadband continuous bremsstrahlung. The drop of spectral brightness around 11 keV might be caused by reabsorption of Cu atoms and ions.

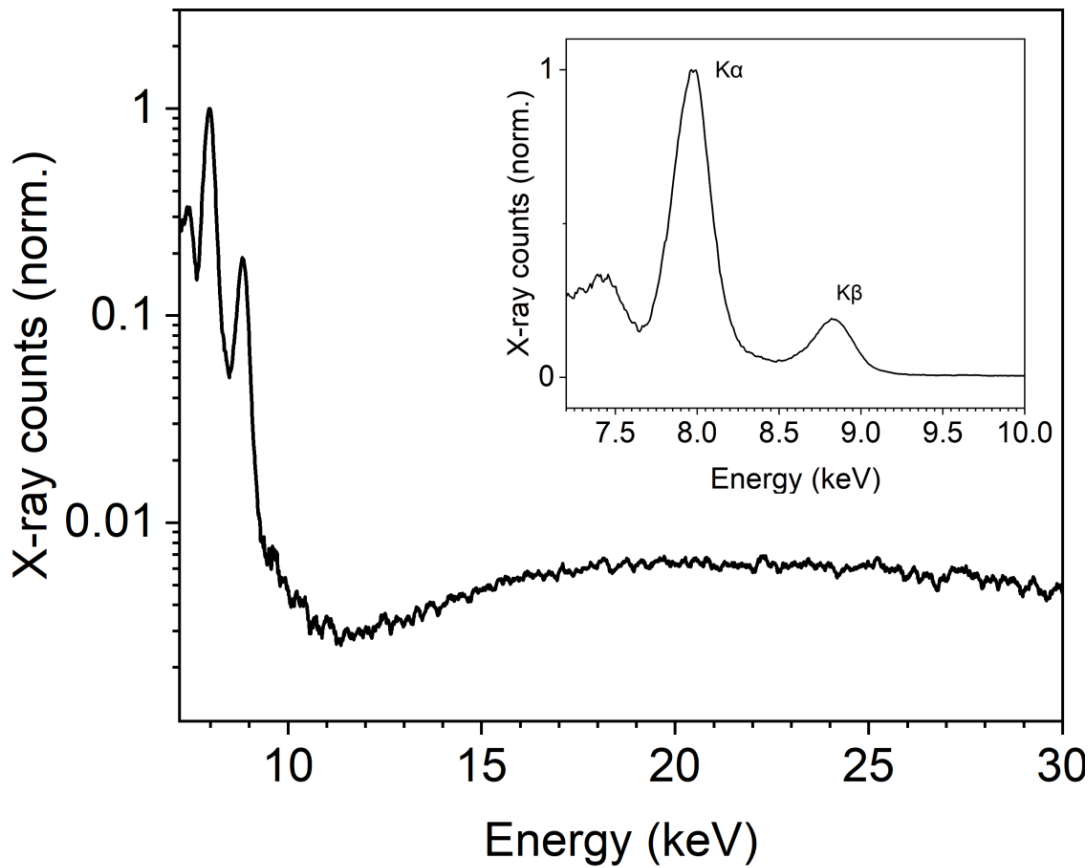

**Figure 3** Emission spectrum of Cu PXS driven by LEGEND laser, depicting characteristic copper lines together with the Bremsstrahlung radiation. Measured with 360 µm thick aluminium filter inserted in front of a single-photon counting silicon detector (AMPTEK Si-PIN detector with thickness of 500 microns). The inset displays the $K_\alpha$ and $K_\beta$ emission lines in higher resolution. The resulting spectrum

is corrected for air absorption, aluminium attenuation, as well as absorption in silicon (Henke *et al.*, 1993). The energy calibration of the spectrometer was done using lines of $^{55}$Fe and $^{241}$Am radioactive sources. The spectrum was measured at distance 1.6 m with an exposure time of 22 minutes.

In the following table, we summarize basic characteristics of the driving lasers and key properties of the X-ray sources driven by each laser. The individual measurements of the X-ray beam are described in the following subsections.

**Table 1** Summary of X-ray source characteristics including driving lasers parameters.

|  |  | LEGEND | L1 ALLEGRA |
| --- | --- | --- | --- |
| Driving laser | Power* | 7.2 W | 14.3 W |
|  | ns contrast | $3\times10^{-3}$ | $< 1\times10^{-7}$ |
|  | 10-ps contrast | $3\times10^{-4}$ | $< 1\times10^{-9}$ |
|  | Intensity** | $1\times10^{17}$ W/cm$^2$ | $5\times10^{17}$ W/cm$^2$ |
|  | Repetition rate | 1 kHz | 1 kHz |
| X-ray source | Plasma source size | ~30 μm | ~30 μm |
|  | Average radiant intensity*** | $10^{11}$ ph/s/4π | $10^{12}$ ph/s/4π |
|  | Montel optics focus size | 100 μm | 100 μm |
|  | Average photon flux in the X-ray focus*** | $2\times10^6$ ph/s | $2\times10^7$ ph/s |

*After the beam transport and beam splitters for pump arm in PXS hutch and diagnostics.

**Considering 10 μm (at 1/e$^2$) focal spot diameter, which was the same for both lasers.

***Only K$_\alpha$ and K$_\beta$ radiation

### 4.2.1. X-ray source size

The size of the X-ray source was measured from a radiograph of perpendicular steel blades with a two-dimensional knife-edge method. The image of the blades with magnification of 7 was recorded onto a back-illuminated X-ray CCD with deep depletion sensor technology with 13.5 μm pixel size (Andor ikon-L 936) providing effective spatial resolution of ~2 μm. The results have shown that the source is symmetric in horizonal and vertical dimension. To assess the X-ray source size, we have applied the '20–80% intensity' criterium and calculated the diameter of the X-ray source to be 30 μm (Figure 4, *left*). Assuming a Gaussian source, the corresponding full width at half maximum (FWHM) source size would be 20 μm.

Such a small source enables PXS to be utilized for X-ray imaging applications. The imaging capabilities of the PXS source are illustrated in Figure 4 *(right)*. For the proof-of-concept demonstration, a bee was placed in front of the exit slit and the transmission image was captured onto a CCD camera with a magnification of 4.3 and exposure time of 60 seconds. Among other possible applications, the near-field ptychography was successfully demonstrated with a compact laboratory plasma source for the first time at PXS beamline (ELI Beamlines) (Fardin *et al.*, 2024).

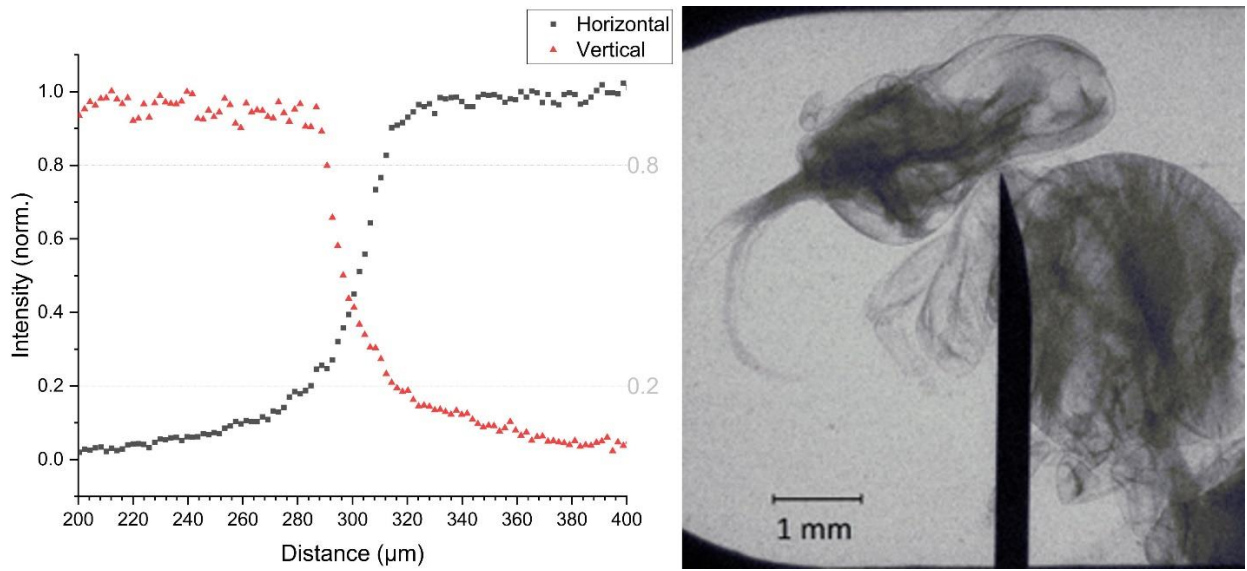

**Figure 4** *(left)* Knife-edge test results measured with LEGEND laser. It shows a normalized intensity profile of a sharp edge of a steel razor blade placed between the source and the detector with ratio of distances being 7. *(right)* Radiography image of a bee acquired with 60 s exposure time. The black pointy object represents a needle fixing the sample.

The post-shot traces on the tape in Figure 5 reveal significantly different interaction regimes of the two driving lasers. While L1 ALLEGRA has produced clean holes with 65 μm diameter, the LEGEND created crater-like traces with a notably asymmetric halo of comparable size. The burnt-through part has only 20 μm in diameter. We believe this is due to different peak intensity and contrast of the two lasers that transfer into different laser-matter interaction scenarios.

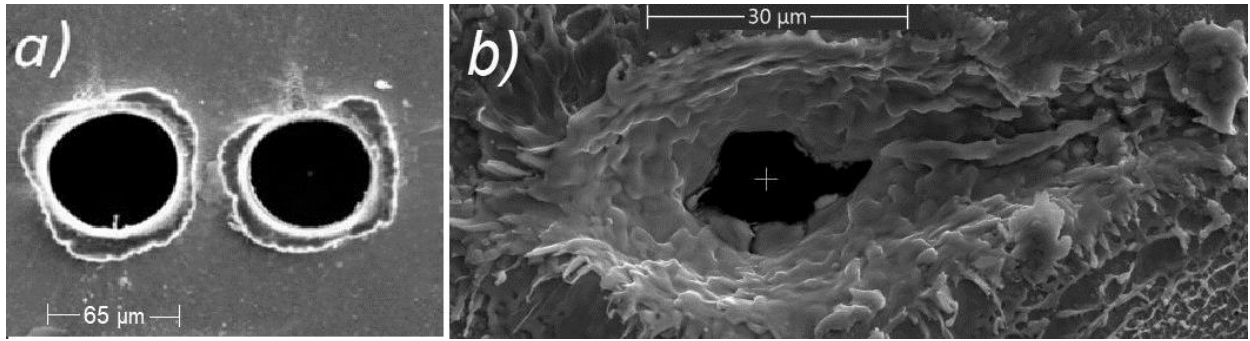

**Figure 5** Comparison of laser-impact for a) L1 ALLEGRA b) LEGEND lasers. The corresponding scale is inserted in each image.

### 4.2.2. Radiant intensity

For estimating the number of photons in each X-ray pulse, listed in Table 1, we utilized the X-ray detector based on PIN diode (xPIN by Rigaku Innovative Technologies Europe). The silicon diode was covered by a 25 μm thick black Kapton window, which blocks all visible light but transmits X-rays. The sensitivity of the diode, with thickness of 300 μm and active area of 7×7 mm$^2$, ranges from 3 to 30 keV peaking at 8 keV. The diode was biased by 50 V to increase its sensitivity and signal-to-noise ratio of the oscilloscope readout. Moreover, the diode can capture each shot at 1 kHz repetition rate to provide reference data for application experiments. For conversion of the photo-charge measured by the diode to the number of photons we used the responsivity of the diode at 8 keV, as most of the photons are emitted in this spectral line (Figure 3).

The radiant intensity of the Bremsstrahlung was estimated from the spectrum at 20 keV. The flux per 1 keV is two orders of magnitude lower in this spectral region compared to the $K_\alpha$ flux.

### 4.2.3. X-ray pulse duration estimate

Laser-driven plasma X-ray source occupies its niche among other X-ray sources mainly for its short pulse duration. Theoretical model predicts sub-picosecond pulses (Reich *et al.*, 2000), that was strongly supported by the indirect measurements(Linde *et al.*, 2001; Zamponi *et al.*, 2009; N. Zhavoronkov *et al.*, 2005). Here, the rapid change in physical properties of the sample was monitored via ultrafast pump-probe scheme. These methods are usually based on monitoring the drop of intensity or shift of the Bragg peak due to the atomic lattice oscillation or observing a drop of reflectivity due to the lattice melting. Successful resolution of the fs dynamics gives an upper limit for the probe pulse duration and is estimated to be 300 – 500 fs. The first cross-correlation measurement of X-ray pulse duration from laser-driven PXS indicated the pulse length to be ~100 fs (Iqbal *et al.*, 2015).

The current understanding of the emission process suggest that the K$_\alpha$ radiation is limited by the laser pulse duration and electron thermalization time (Rousse, Antonetti *et al.*, 1994; Kieffer *et al.*, 1996; Teubner *et al.*, 1996). To the best of the authors' knowledge, direct measurements of sub-picosecond pulses in the hard X-ray range have not yet been performed. However, continued advancements in ultrafast X-ray detectors are driving progress toward achieving 100-fs streak camera technology (Qiang *et al.*, 2009; Liu *et al.*, 2003; Toufexis & Dolgashev, 2019; Li *et al.*, 2015; Feng *et al.*, 2010), bringing us closer to the PXS pulse duration direct measurements. Recently, an indirect measurement technique was suggested predicting the achievable temporal resolution to be few tens of femtoseconds (Nazarkin *et al.*, 2004).

The time-resolved structure of the accompanying Bremsstrahlung radiation is an important parameter for the time-resolved absorption spectroscopy. The pulse durations of both Bremsstrahlung and K$_\alpha$ emission were modelled, suggesting similar pulse duration (Reich *et al.*, 2007).

### 5. Experimental end-stations

In addition to its compact footprint, a key feature of our PXS is its ability to produce pulses with durations of the order of hundreds of femtoseconds. To fully utilize this capability, we have identified two time-resolved applications - X-ray diffraction and X-ray spectroscopy - that are ideal for studying ultrafast sample dynamics initiated by photoexcitation in a pump-probe setup. For this arrangement, a pump beam is generated by splitting 10% of the main driving laser beam earlier in the beam transport section and synchronized with the X-ray pulse using a delay line. The pump pulse, typically at 800 nm, can be converted to higher harmonics or used to generate a supercontinuum to optimize the sample response.

### 5.1. The X-ray diffraction end-station

The TREX (Time Resolved Experiments with X-rays) end-station at ELI Beamlines is a state-of-the-art instrument designed for time-resolved X-ray diffraction and scattering experiments to study fast processes in crystallographic samples. The end-station is a versatile diffractometer designed to investigate electronic and structural dynamics in a wide range of scientific fields, including chemical reactions, biomolecular dynamics, and material science. The setup features a Eulerian cradle diffractometer (STADIVARI from STOE & Cie GmbH) that enables precise sample positioning through 3-axis orthogonal translations and three rotations (SmarAct, GmbH). This configuration supports a wide range of experiments, including powder diffraction, wide-angle scattering, and studies on thin solid film samples. The main detector is a Jungfrau detector (3 panels, total 1.5 megapixel with 75 μm pixel size) from PSI, Switzerland (Mozzanica *et al.,* 2018; Leonarski *et al.,* 2018), which can be translated 40-400 mm from the sample and rotated in the

θ-2θ mode. An additional X-ray back-illuminated charge-coupled device with deep depletion chip technology (Andor ikon-L 936) is available for alignment purposes. Some experiments were performed using a hybrid photon counting detector Eiger X 1M (DECTRIS). For single crystal protein diffraction experiments, sample cooling can be provided by a cryogenic gas stream (Oxford Cryosystems).

Femtosecond X-ray pulses generated by the PXS are conditioned using a motorized Montel optics ((Montel, 1957) AXO Dresden GmbH) for monochromatization, focusing, and collimation (Shymanovich *et al.*, 2008). The 15 cm-long Montel mirror with acceptance angle of 16.8 mrad is placed 11.5 cm from the source delivering $K_\alpha$ radiation onto a probed sample with enhanced reflectivity thanks to a Ni/C multilayer coating. The focus is 54 cm from the mirror's exit edge. Considering the primary focal length of 190 mm and secondary one of 615 mm, the magnification of the optics is approximately 3.2, resulting in a divergence of the focused beam to be 5 mrad. The beam intensity depends on the driving laser and for L1 ALLEGRA laser operation was measured to be $\sim 2 \times 10^7$ photons/s with a focal spot size of ~100 μm.

The Jungfrau detector, a key component of the TREX end-station, operates at a high repetition rate of 1 kHz, generating a substantial amount of data that can fill several terabytes of disk space daily. To tackle this data challenge, ELI Beamlines is developing customized software solutions for calibrating and converting the raw data into a common photon-per-pixel format. Simultaneously, progress is being made in standardizing 2D image formats based on the HDF5 specification, which will promote harmonization with other large-scale facilities and enable more efficient data sharing and collaboration among researchers.

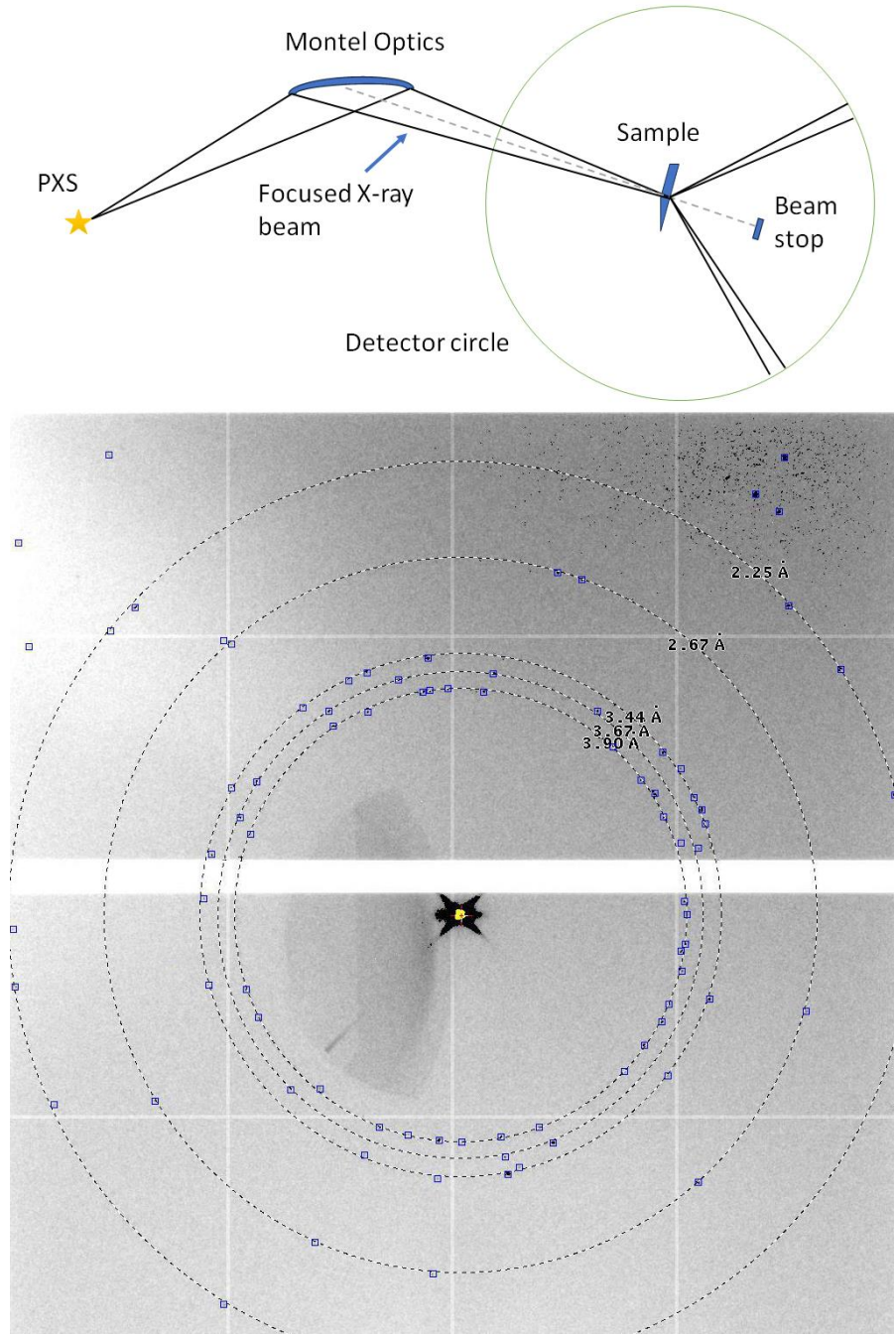

**Figure 6** *(top)* Scheme of the TREX end-station. Scheme of 1D parallel beam optics geometry with fixed entrance and exit apertures of Montel optics ASTIX++ from AXO Dresden. *(bottom)* Diffraction from hexagonal and cubic ice formed at 100 K. Measured with DECTRIS Eiger X 1M at distance 46.4 mm from the sample. Localization of diffraction peaks was performed using ADXV with a signal to noise ratio threshold of 7. The shadow of the direct beam is visible on the left side of the image

owing to the absence of the collimator. Diffraction rings of the ice are at interplanar spacing of 2.25 Å, 2.67 Å, 3.44 Å, 3.67 Å, and 3.90 Å.

The station's modular design allows for adaptations to accommodate user-initiated experiments, providing researchers also with the flexibility to install their own equipment. The combination of femtosecond X-ray pulses and advanced detectors at the TREX end-station will enable high-resolution time-resolved studies of various phenomena, such as phase transitions, chemical reactions, and protein dynamics (Schmidt, 2019; Rischel *et al.*, 1997). Future plans include the implementation of wide-angle X-ray scattering (WAXS), thin film reflectometry, and surface diffraction techniques.

**5.2. The X-ray spectroscopy end-station**

X-ray spectroscopy is an experimental tool that gives access to information about the electronic structure of materials. X-ray emission spectroscopy (XES) allows to probe occupied states while X-ray absorption spectroscopy (XAS) traces the difference between the incoming and outgoing X-ray signal revealing the photon fraction absorbed in the sample and thus exploring unoccupied states. Fundamentally, the two techniques provide complementary information. Thus, the goal of the X-ray spectroscopy station at ELI Beamlines is simultaneous acquisition of emission and absorption data via development of parallel measurements accommodated around the Cu-tape PXS with sub-ps temporal resolution.

Typically, an X-ray spectroscopy experiment consists of an X-ray source, the sample under study (with corresponding environment if required) and a spectrometer. The latter can be a single-body detector, e.g. of a microcalorimeter type (Doriese *et al.*, 2017), in a very compact setup, or a crystal-based solution (Sokaras *et al.*, 2013) that can achieve much higher spectral resolution. In the case of a crystal-based design, there are two alternative approaches to spectrum collection. In Johann and Johansson-type spectrometers, a 2D focusing crystal selects a single energy and focuses the X-rays down to a point. To acquire the full spectrum an energy scan is needed, which requires moving the crystal and the detector. Alternatively, in von Hamos geometry (Szlachetko *et al.*, 2012), a crystal bent in one dimension and energy resolving in another, allows for collection of full spectra simultaneously, eliminating the need for scanning, but reducing the solid angle of collection for given photon energy. For a focusing crystal few alternatives exist, such as segmented (Németh *et al.*, 2016) or continuous Si wafers glued to a cylindrically bent substrate, Highly Oriented Pyrolytic Graphite (HOPG) (Legall *et al.*, 2006) and Highly Annealed Pyrolytic Graphite (HAPG) (Malzer *et al.*, 2021).

A requirement to resolve chemical changes of the oxidation state (i.e. well below 1 eV) has imposed the choice of a crystal-based spectrometer. However, the space constraints in the experimental hutch called for a strategic selection of a compact design. Thus, we have selected a von Hamos type spectrometer (Zymaková *et al.*, 2023). An additional benefit from this choice is that it enables us to use the spectrometer solutions developed for the PXS at other X-ray sources available at ELI Beamlines such as plasma betatron X-ray source (Chaulagain *et al.*, 2022) providing a user with wide choice of source parameters.

A polycapillary X-ray focusing optics has been incorporated (Zymaková *et al.*, 2020) for experiments with inhomogeneous samples and lower requirements on temporal resolution. The system can be used for solid, powder or liquid samples. The ability to handle liquid samples paves the way for biological sample studies. A few different liquid sample delivery systems, such as a wire-guided jet (Picchiotti *et al.*, 2023), colliding jet (Koralek *et al.*, 2018) and microliter stirred cell (Fanselow *et al.*, 2022) are available for users at ELI Beamlines.

## 6. Discussion and outlook

### 6.1. Comparison to other laser-driven sources and outlook

Among established PXS beamlines, our source radiant intensity reaches a current best-efforts flux for such sources, ranging up to $10^{11}$–$10^{12}$ ph/s/4π (Zhao *et al.*, 2022; Holtz *et al.*, 2017). An exceptional flux of $10^{12}$ ph/s is reached by driving laser with pulse energy as low as 3 mJ thanks to the use of a long 5 μm wavelength (Koç *et al.*, 2021). An additional X-ray yield can be reached by optimizing a pre-pulse (Lu *et al.*, 2024), reaching $10^5$ ph/sh on the sample. Similar flux values were achieved by (Afshari *et al.*, 2020), utilizing high-energy laser pulses of 100 mJ.

Depending on the choice of the focusing optics, an average loss at monochromatization is four (Zhao *et al.*, 2022), five (Koç *et al.*, 2021) or six orders of magnitude (Holtz *et al.*, 2017; Schick *et al.*, 2012). Our Montel optics, which belongs to the longer-focus category, the flux loss is $10^6$ orders of magnitude compared to the full solid angle.

Our radiant intensity will be further enhanced through improved stability achieved by a redesigned interaction chamber. The current distance between the guiding bearings and the plasma is insufficient, leading to gradual damage of the bearings and consequent tape quivering. Furthermore, an active beam stabilization that utilizes quadrupole diode for monitoring the laser beam position and pointing will prevent the source from drifting due to temperature drift of the system.

### 6.2. PXS as an alternative to large-scale facilities

Time-resolved diffraction as well as X-ray spectroscopy are well-established techniques in the large-scale facilities, such as synchrotrons and XFELs. While ultrafast high-intensity pulses from XFEL ensure exceptionally fast data collection, 'diffraction before destruction' is assumed (Chapman *et al.*, 2011). Nevertheless, as $10^{12}$ photons are delivered to the sample in matter of just 1-100 fs (R. Abela *et al.*, 2017), it raises the moderate concern on the possibility of rapid changes in protein structure in response to extreme irradiation, especially where pulse durations are greater than a few tens of femtoseconds (Nass, 2019; Nass *et al.*, 2020; Garman & Weik, 2023; Brinkmann & Hub, 2016; Arnlund *et al.*, 2014; Ansari *et al.*, 1985). In addition, the destruction of a crystal after a single exposure from a high intensity XFEL pulse results in high rates of sample consumption.

Synchrotrons provide flux up to $10^{15}$ photons per second (Garman & Weik, 2023) but usually operate in a pseudo- continuous wave mode, limiting the achievable temporal resolution when in normal operation. Detector gating (Donath *et al.*, 2023), femtoslicing (Labat *et al.*, 2018) and rotating choppers may be used (Meents *et al.*, 2009; Cammarata *et al.*, 2009), however, radiation damage remains an important consideration.

Both XFEL and Synchrotron demand large infrastructure for their operation. Thus, plasma X-ray sources emerge as a promising alternative beamline for time-resolved experiments. PXS combines the advantages of a pulsed source and short pulse duration. The plasma X-ray source is a tabletop and potentially could fit into a small university lab. Their relatively low cost makes them an affordable option. However, the downside of PXS is its unstable flux and comparatively low brilliance. The latter can usually be overcome by stroboscopic (Kim *et al.*, 2002) or multiplexing (Klureza *et al.*, 2024) approach, where the ensemble of the sample responses is gathered and merged together. This makes PXS a suitable tool for studying ultrafast reversible processes, such as superlattice oscillations, X-ray absorption spectroscopy, etc.

## 7. Conclusion

In this paper, we have introduced the capabilities of a newly commissioned 1 kHz laser-driven Cu-tape plasma X-ray source, showcasing its versatility and potential for a range of scientific applications. The source can be driven by either a sub-20 fs near-infrared OPCPA system or a commercial Ti:Sapphire laser. Our findings highlight the differences in X-ray beam characteristics and photon flux achievable with different laser systems. A significantly higher X-ray yield with the OPCPA driver is given by contribution of several factors. First, higher laser pulse energy and shorter pulse duration lead to higher laser intensity on target. In addition, better ns contrast further enhances the $K_\alpha$ yield as compared to the case driven by conventional Ti:Sapphire laser.

The ability to deliver photon fluxes up to the order of $10^{12}$ $K_\alpha$ photons/$4\pi$/s underlines the high performance of our X-ray source. Additionally, we have outlined the potential of this system for ultrafast X-ray diffraction and X-ray absorption and/or emission spectroscopy, paving the way for innovative research studying ultrafast structural dynamics. As the system is now available for user-based access, we anticipate it will serve as a powerful tool for the scientific community, enabling a wide range of experiments in the fields of material science, chemistry, biology, and physics.

**Acknowledgements**

We express gratitude to Martin Přeček for the electron microscope images of the used tape. This research was supported by the project Advanced research using high intensity laser produced photons and particles (CZ.02.1.01/0.0/0.0/16_019/0000789) from European Regional Development Fund (ADONIS). Y.P and I.B. thank the Charles University Grant agency, project GA UK No. 113-10/252615. Y.P. thanks Charles University grant SVV–2024–260720 and IMPULSE – European Regional Development Fund (871161 — IMPULSE — H2020-INFRADEV-2018-2020 / H2020-INFRADEV-2019-1). S.C. and L.F. are supported by the EPSRC New Investigator Award EP/X020657/1 and Royal Society RGS/R1/231027.

**Conflicts of interest**   Authors declare that there are no conflicts of interest.

**Data availability**   Data supporting this publication are available upon reasonable request to the corresponding author.